\documentclass[aps,twocolumn,preprintnumbers,nofootinbib,superscriptaddress]{revtex4-1}
\usepackage{amsmath} \usepackage{graphicx} \usepackage{amsfonts}
\usepackage{array} \usepackage{amsthm} \usepackage{bm}
\usepackage{palatino} \usepackage{mathpazo} 
\usepackage{supertabular}

\usepackage[breaklinks]{hyperref}
\usepackage{color}

\newcommand{\be}{\begin{equation}}
\newcommand{\ee}{\end{equation}}
\newcommand{\ba}{\begin{eqnarray}}
\newcommand{\ea}{\end{eqnarray}}
\newcommand{\bal}{\begin{align}}
\newcommand{\eal}{\end{align}}

\newcommand{\dd}{\text{d}}

\newcommand{\bw}{\begin{widetext}}
\newcommand{\ew}{\end{widetext}}

\begin{document}


\title{Comment on {\textgravedbl}Regularized big bang singularity{\textacutedbl}}

\author{Mustapha Azreg-A\"{\i}nou}
\affiliation{Ba\c{s}kent University, Engineering Faculty, Ba\u{g}l{\i}ca Campus, 06780-Ankara, T\"{u}rkiye}


\begin{abstract}
We show that the regularized-big-bang-singularity model discussed in Phys. Rev. D \textbf{100}, 023536 (2019), suffers from the same drawbacks of the Friedmann-Lema\^{\i}tre-Robertson-Walker (FLRW) universe, in that the $T$-odd solution has the same divergences at the initial new time $T$ as the FLRW universe, while the $T$-even solution is exempt from such divergences.
\end{abstract}

\maketitle

\section*{}
A tentative to regularize the initial singularity at $t=0$, of the standard Friedmann-Lema\^{\i}tre-Robertson-Walker (FLRW) flat universe, was made in Ref.~\cite{RBBS} upon replacing the standard cosmic time coordinate $t > 0$ by an extended time $\tau\in \mathbb{R}$, then removing from the real $\tau$-axis the open interval $(-b,b)$. The antipodal points $\tau=-b$ and $\tau=b$ were identified and a new cosmic time coordinate $T$ was defined by the following two-piecewise function
\begin{equation}\label{eq1}
T=\begin{cases}
	\sqrt[4]{\tau^4-b^4} & \text{for }\tau \geq b\\
	-\sqrt[4]{\tau^4-b^4} & \text{for }\tau \leq -b
\end{cases}\,.
\end{equation}

Such a topological surgery does not guaranty analyticity as we shall see shortly. In terms of $T$, the so-called modified FLRW line element takes the form~\cite{RBBS}
\begin{align}
\label{eq20}\dd s^2=&-c^2\,f(T)\,\dd T^2 + a^2(T)\delta_{kl}\dd x^k	\dd x^l\,,\,,\\
\label{eq2}\quad=&-c^2\,\frac{T^6}{(T^4+b^4)^{3/2}}\,\dd T^2 + a^2(T)\delta_{kl}\dd x^k	\dd x^l\,,
\end{align}
where $c$ is the speed of light and $a(T)=a(\tau(T))$ with $\tau(T)$ being the two-piecewise inverse function of $T(\tau)$~\eqref{eq1}, given by~\cite{RBBS}
\begin{equation}\label{eq3}
\tau(T)=\begin{cases}
	\sqrt[4]{T^4+b^4} & \text{for }T \geq 0\\
	-\sqrt[4]{T^4+b^4} & \text{for }T \leq 0
\end{cases}\,.
\end{equation}

A two-piecewise expression for $a(T)$, the so-called $T$-odd solution, was derived in~\cite{RBBS} in the case where the energy-momentum tensor is that of a homogeneous perfect fluid with the equation of state: pressure $=c^2$ (energy density)/3 [$P(T)=c^2\rho(T)/3$]:
\begin{equation}\label{eq5}
a(T)=\begin{cases}
	\sqrt[8]{(T^4+b^4)/(T_0^4+b^4)} & \text{for }T > 0\\
	-\sqrt[8]{(T^4+b^4)/(T_0^4+b^4)} & \text{for }T \leq 0
\end{cases}\,.
\end{equation}
This expression of $a(T)$ is defined at $T=0$ and has a finite jump discontinuity there: it jumps by $2\sqrt[8]{b^4/(T_0^4+b^4)}$ from $-\sqrt[8]{b^4/(T_0^4+b^4)}$ at $T=0$ to $\sqrt[8]{b^4/(T_0^4+b^4)}$ at $T=0^+$. This jump in the value of the scale factor $a(T)$ yields a Dirac-delta behavior of $a'(T)\equiv \dd a(T)/\dd T$, as we shall see shortly. Since $a'(T)$ is related to $\rho(T)$ by the field equations [Eqs.~(3.3) of~\cite{RBBS}]:
\begin{equation}\label{field}
\Big(1+\frac{b^4}{T^4}\Big)^{3/2}\Big(\frac{1}{a(T)}\frac{\dd a(T)}{\dd T}\Big)^2=\frac{8\pi G}{3}~\rho(T)\,,
\end{equation}
$\rho(T)$ also will have a Dirac-delta behavior at $T=0$. However, a claim was made in Ref.~\cite{RBBS} that $a(T)$~\eqref{eq5} {\textquotedblleft has continuous first-, second-, and third-order derivatives at $T = 0$ and a discontinuous fourth-order derivative at $T = 0$ ...\textquotedblright}. A function that has a jump discontinuity at some point, it does not possess a first-, or any order-, derivative at that point~\cite{Thomas}. 

For instance, using~\eqref{eq2} we obtain the Ricci curvature scalar $R$ by
\begin{align}\label{eq4}
&R(T)=\frac{6 \sqrt{T^4+b^4}}{c^2 T^7 a(T)^2}\times\\ 
&\{T (T^4+b^4) a'(T)^2+a(T) [-3 b^4 a'(T)+T (T^4+b^4) a''(T)]\},\nonumber	
\end{align}
where $a''(T)\equiv \dd ^2 a(T)/\dd T^2$, and this has a Dirac delta at $T=0$.

Introducing the unit step function $u(T)$ defined by~\cite{Thomas,Roper}
\begin{equation}\label{eq6}
u(T)=\begin{cases}
	1 & \text{for }T \geq 0\\
	0 & \text{for }T < 0
\end{cases}\,,
\end{equation}
we can write~\eqref{eq5} as
\begin{equation}\label{eq7}
a(T)=[1-2u(-T)]\sqrt[8]{(T^4+b^4)/(T_0^4+b^4)}\,.
\end{equation}
With $\dd u(T)/\dd T=\delta(T)$ (Dirac delta), we obtain\footnote{In obtaining the first term in $a'(T)$ we have used the rule $f(T)\delta(T)=f(0)\delta(T)$, where $f(T)$ is some regular function at $T=0$ (no pole at $T=0$). We have used the same rule along with $\text{const}\times \delta'(T)=0$ to obtain $a''(T)$.}
\begin{align}
&a'(T)=2\Big(\frac{b^4}{T_0^4+b^4}\Big)^{1/8}\delta(T) + \frac{T^3a(T)}{2(T^4+b^4)}\,,\nonumber\\
&a''(T)=-\frac{T^2(T^4-6b^4)a(T)}{4(T^4+b^4)^2}\,,\nonumber\\
\label{eq8}&\rho(T) =\frac{3}{32 \pi  G \sqrt{T^4+b^4}} + \frac{3\sqrt{b}\sqrt{(T^4+b^4}}{4\pi GT^3(T_0^4+b^4)^{1/8}a(T)}~\delta (T),
\end{align}
where products of Dirac delta functions have been removed. Setting $\rho_0\equiv 3/\big(32\pi G\sqrt{T_0^4+b^4}\,\big)$, the first term in~\eqref{eq8} takes the form $\rho_0\sqrt{(T^4+b^4)/(T_0^4+b^4)}$, which is the same term in Eqs.~(3.5) of~\cite{RBBS}. The pressure has a similar expression to~\eqref{eq8} and only the first terms in the expressions of $P(T)$ and $\rho(T)$ (not including the Dirac delta) obey the relation $P(T)=c^2\rho(T)/3$.

The singularity in the expressions of $R(T)$ and $\rho(T)$ is real and appears not just because of the Dirac delta but also because of the presence in their denominators of powers of $T$. This is important to justify the jump in the value of the scale factor $a(T)$ at $T=0$. This {\textgravedbl}super inflationary{\textacutedbl}, still speculative, phase in the evolution of the universe cannot be justified if pressure and density remain finite at $T=0$.

In Ref.~\cite{RBBS} the dynamical variable was assumed to be $A(T)=a^2(T)$, which does not have a jump discontinuity at $T=0$, and this yields $\rho(T)\propto A'(T)^2/A(T)^2$ producing the first term in~\eqref{eq8} without Dirac delta since $A'(T)$ is regular at $T=0$. We emphasize here that $A(T)$ does not represent a physical or a geometric entity and, based on the field equations~\eqref{field}, any abrupt change in the value of the scale factor $a(T)$, produces a divergence in the value of $\rho(T)$.

Another metric ansatz, similar to~\eqref{eq20}, was proposed in~\cite{rev,rev2} with $f(T)$ [different from that shown in~\eqref{eq2}] was chosen of the form
\begin{equation}\label{eqp1}
f(T)=\frac{T^2}{T^2+b^2}\,,
\end{equation}
by which the so-called $T$-even solution was derived in~\cite{RBBS,rev}
\begin{equation}\label{eqp2}
a(T)=\sqrt[4]{(T^2+b^2)/(T_0^2+b^2)}\quad  \text{for }T \in \mathbb{R}\,,
\end{equation}
and this does not suffer from the same drawbacks of the $T$-odd solution.

A final comment is in order. It is possible to show that (and this is out of the scope of the present comment~\cite{equiv}), for any $f(T)$ and for a spatially flat homogeneous Friedmann-Lema\^{\i}tre-Robertson-Walker universe with no cosmological constant, the regularized-big-bang-singularity model is equivalent to the gravitational-particle production model initiated by Prigogine et. al~\cite{GPPP1,GPPP2,GPPP3}.





\begin{thebibliography}{99}
\bibitem{RBBS}F.R. Klinkhamer,
\href{https://doi.org/10.1103/PhysRevD.100.023536}{Phys. Rev. D \textbf{100}, 023536 (2019)}, arXiv:1903.10450

\bibitem{Thomas}J. Hass, C. Heil, P. Bogacki, M.D. Weir,
\textit{Thomas' Calculus}, 15th Edition, Pearson Education, Boston (2023)

\bibitem{Roper}L.D. Roper,
\textit{The Dirac Delta Function in Physics}, Physical Biological Sciences Misc, Virginia (1972)

\bibitem{rev}F.R. Klinkhamer,
\href{https://arxiv.org/abs/2604.00077}{Proceedings of the Corfu Summer Institute 2025}, {\textgravedbl}School and Workshops on Elementary Particle Physics and Gravity{\textacutedbl}, Corfu, Greece, arXiv:2604.00077

\bibitem{rev2}F.R. Klinkhamer,
\href{https://doi.org/10.1142/S0217732325300101}{Mod. Phys. Lett. A \textbf{40}, 2530010 (2025)}, arXiv:2412.03538

\bibitem{equiv}M. Azreg-A\"{\i}nou,
\textit{Equivalence of the gravitational-particle-production and regularized-big-bang-singularity models}, submitted to a journal

\bibitem{GPPP1}I. Prigogine, J. Geheniau, E. Gunzig, P. Nardone,
\href{https://doi.org/10.1073/pnas.85.20.7428}{Proc. Natl. Acad. Sci. U.S.A. \textbf{85}, 7428 (1988)}; \\
\href{https://doi.org/10.1007/BF00758981}{Gen. Relativ. Gravit. \textbf{21}, 767 (1989)}

\bibitem{GPPP2}J. A. S. Lima and I. Baranov,
\href{https://doi.org/10.1103/PhysRevD.90.043515}{Phys. Rev. D \textbf{90}, 043515 (2014)}, arXiv:1411.6589

\bibitem{GPPP3}N. Komatsu,
\href{https://arxiv.org/abs/2512.19911}{arXiv:2512.19911},




\end{thebibliography}
\end{document}